\documentclass[12pt,preprint]{aastex}

\slugcomment{Accepted by The Astronomical Journal}

\shorttitle{Water masers in Bok globules}
\shortauthors{de Gregorio-Monsalvo et al.}

\begin{document}


\title{High-resolution observations of water masers in Bok globules}


\author{Itziar de Gregorio-Monsalvo\altaffilmark{1}, Jos\'e
  F. G\'omez\altaffilmark{2}, Olga Su\'arez\altaffilmark{1}, 
   Thomas B. H. Kuiper\altaffilmark{3}, Guillem
  Anglada\altaffilmark{2}, Nimesh A. Patel\altaffilmark{4},
  Jos\'e M. Torrelles\altaffilmark{5}} 

\altaffiltext{1}{Laboratorio de Astrof\'{\i}sica Espacial y F\'{\i}sica
  Fundamental (INTA), Apartado 50727, E-28080 Madrid, Spain;
  itziar@laeff.inta.es, olga@laeff.inta.es}
\altaffiltext{2}{Instituto de Astrof\'{\i}sica de Andaluc\'{\i}a
  (CSIC), Apartado 3004, E-18080 Granada, Spain; jfg@iaa.es, guillem@iaa.es.}
\altaffiltext{3}{Jet Propulsion Laboratory, California Institute of
  Technology, 4800 Oak Grove Drive, Pasadena, CA 91109, USA; kuiper@jpl.nasa.gov.}
\altaffiltext{4}{Harvard-Smithsonian Center for Astrophysics, 60
  Garden Street, Cambridge, MA 02138, USA; \\ npatel@cfa.harvard.edu.}
\altaffiltext{5}{Instituto de Ciencias del Espacio (CSIC) and Institut d'Estudis Espacials de Catalunya, Facultat de F\'{\i}sica, Planta 7,
Universitat de Barcelona, Av. Diagonal 647, E-08028 Barcelona, Spain; torrelles@ieec.fcr.es.}


\begin{abstract}
We present Very Large Array observations 
at 1.3 cm of several water maser detections obtained by previous
single-dish studies of Bok globules in the Clemens \& Barvainis (1988;
CB) catalog. 
We report water maser emission in CB 3 (CB3-mm), CB 54 (IRAS
07020$-$1618), CB 101
(IRAS 17503$-$0833), and CB 232 (IRAS 21352+4307), and non-detection
towards CB 65 (IRAS 16277$-$2332) and CB 205 (IRAS 19433+2743).  These
are the first reported interferometric observations of water masers in
Bok globules of the CB catalog.
We also present single-dish observations of millimeter and
centimeter spectral lines  towards CB 101 (IRAS 17503$-$0833) and CB
65 (IRAS 16277$-$2332).
All the maser emission seems to be associated with star forming
regions hosting 
bipolar molecular outflows, except IRAS 17503$-$0833 in CB 101, 
which we suggest to be a possible Mira evolved star, and 
IRAS 16277$-$2332 in CB 65, of unknown nature. 
We have used the precise position of the maser emission to derive
 information about the powering source of the masers. By analyzing the
spatio-kinematical distribution
of the water masers, we confirm the millimeter source CB 3-mm as the
most likely powering source of the CB 3 masers. 
We propose the near-IR source CB 232 YC1-I as the best candidate for 
pumping the maser emission observed in CB 232, while in CB 54, we suggest that
the pumping source of the masers could be located 
at the position of an elongated feature observed in near-infrared maps.

\end{abstract}





\section{Introduction}

Bok globules \citep{Bok47} are small ($\la 1$ pc), isolated clouds, some of which are sites of low- and intermediate-mass star
formation \citep{Yun90}. They have been observed to be in different
evolutionary stages, 
from starless dark cores \citep{Yun96,Kan97} to sites hosting T-Tauri stars 
\citep{Yun95,Lau97}.  These globules have traditionally been considered
important laboratories for the study of star formation processes,
since their small size and relative simplicity make these processes
less likely to be affected by confusion from different generations of
young stellar objects. Another important aspect of star formation in
Bok globules is that it may explain the somewhat puzzling existence of
pre-main-sequence objects apparently not related with
known molecular clouds. At least some of these objects may have
originated in already dispersed Bok globules \citep{Lau97}.

Among the observational tools used in star-formation studies, water
masers have proved to be a very powerful one.
Water maser emission at 22 GHz is a good tracer of mass-loss activity
in young stars \citep{Rod80,Fel92,Xia95,Deb05}. In the low-mass 
young stellar objects (YSOs) this signpost of mass loss phenomena
occurs in their most embedded phase, at the earliest stages of evolution. 
This period is characterized by the presence
of the disk/young stellar object/outflow system 
\citep{Shu87} and a large amount of circumstellar dust. 
In the low-mass YSOs framework, the youngest evolutionary stage is
represented by Class 0 sources \citep{And93}, which show the most
powerful and collimated molecular outflows (e.g., \citealt{Hir06}, and
references therein), centimeter emission (e.g. \citealt{Ang95}), and host
 strong submillimeter and millimeter continuum emission \citep{And93,And94}. 

Water maser studies have shown that this emission provides a good
characterization of the age of low-mass YSOs (e.g. \citealt{Fur01}), with
Class 0 sources being the most probable candidates to harbor water maser
emission, due to the interaction of powerful jets with a
larger amount of circumstellar material. This fact makes sources
that host water maser emission good candidates for being in a very
early stage of its evolution. In addition, several water maser surveys 
towards low-mass YSOs have shown that this emission tends to be
located close (within several hundred AU) to the central powering source
\citep{Che95, Cla98, Fur00, Fur03}. These characteristics make the
water maser phenomenon well-suited to derive information about the 
location of the exciting source of the mass-loss phenomena observed 
in young star forming regions.  
Moreover, maser emission can be a useful tool for studying the physical
conditions and kinematics of the gas surrounding the most embedded YSOs 
at very high angular resolution ($\leq$ 1 mas), due to the very specific conditions which pump the maser emission (hundreds of K
and n$_{\rm H_{2}}\simeq 10^{8}-10^{10}$ cm$^{-3}$) \citep{Cla98,Fur00,Pat00,Tor03,God05,Mar05,Vle06}.
These conditions can be generated in shocked gas compressed by
winds \citep{Che95,Tor97,Fur00,Mos00}, as well as in circumstellar disks 
\citep{Fie96,Tor96,Tor98,Set02,Bra03}. This dichotomy has also been
suggested to be a possible evolutionary sequence, with water masers tracing
gravitationally bound material (e.g., circumstellar disks) in the
youngest sources, and outflows in more evolved YSOs \citep{Tor97,Tor98}. 
Recently, the hot dense infalling gas after the accretion shock has
also been proposed as a good environment for pumping the maser emission \citep{Men04}.

Therefore, there are different evolutionary aspects related to water maser
emission (e.g., the likely occurrence of masers in Class 0 YSOs,
and their association with either disks and jets). Bok globules are appropriate sites to study
evolution of YSOs, since they are
usually identified as dark patches in optical images, without any
selection criterion related to possible star-forming activity
(\citealt{Cle88}, hereafter CB; \citealt{Bou95}), 
and therefore they may span a large range of
evolutionary stages. Surprisingly, studies of water masers in Bok
globules are relatively scarce. Recently, our group undertook a
sensitive and systematic single-dish survey for water masers in 207 
positions within Bok globules \citep{Gom06}, using NASA's 70m antenna 
in Robledo de Chavela (Spain), which provided six new detections in
\objectname{CB 34}, \objectname{CB 54}, \objectname{CB 65},
\objectname{CB 101}, \objectname{CB 199}, and \objectname{CB 232}.
Before this survey, only \citet{Scap91} conducted a 
search for water masers 
specifically in Bok globules, with 80 target positions and only one
detection towards \objectname{CB 3}. Other
surveys for water masers included some Bok globules within their
target positions \citep[e.g.,][]{Fel92,Pal93,Wou93,Per94,Cod95}, although
they did not find any detection in these types of sources.

Using as a reference the catalogs of Bok globules of 
CB in the northern hemisphere, and \citet{Bou95} in the southern one
(probably the more complete ones available), 
we are only aware of  9 Bok globules in those catalogs that are 
apparently associated with water
maser emission: The six new detections mentioned above in the survey
by \citet{Gom06}, the detection towards CB 3 reported by
\citet{Scap91}, as well as the detections obtained by \citet{sch75}
and \citet{Nec85} in  \objectname{CB 39} and \objectname{CB 205}
respectively. All these reported detections of masers in Bok globules 
present single-dish data. However, interferometric high-angular 
resolution observations are
necessary to pinpoint accurately the actual pumping source of water
maser emission among several candidates, and to determine whether
maser emission in these Bok globules tends to trace collimated jets or
circumstellar disks. 

In this paper we present for the first time interferometric,
high-angular resolution observations using the Very Large Array (VLA) of the National
Radio Astronomy Observatory\footnote{The National Radio Astronomy
  Observatory is a facility of the National Science Foundation
  operated under cooperative agreement by Associated Universities,
  Inc.} of some of these water masers in Bok globules. 
This paper is structured as follows: in \S 2 we describe our observations
and data processing; in \S 3 we present our observational
results and we discuss them; we summarize our conclusions in \S 4.

\section{Observations and data processing}
\subsection{Target selection}
We present interferometric observations using the VLA in the Bok
globules in which water masers were 
detected in the survey by \citet{Gom06} between 2002 April
and  2003 May (i.e., CB 54, CB 65, CB 101, and CB 232), 
as well as CB 3 \citep{Scap91} and CB 205 \citep{Nec85}. 
The interferometric data have been
complemented with a multimolecular single-dish study at millimeter
and centimeter wavelengths using the IRAM-30m and Robledo-70m
antenna of CB 101 (IRAS 17503$-$0833) and CB 65
(IRAS 16277$-$2332), in order to better understand the nature of these
sources, and determine the internal structure of their
surrounding region.

\subsection{VLA Observations}
We observed simultaneously the 6$_{16}$-5$_{23}$ transition 
of H$_{2}$O (rest frequency = 22235.080 MHz) and continuum at 1.3 cm with the VLA, toward 
the Bok globules CB 54, CB 65, CB 101, CB 205, and CB 232. 
The observations were carried out on 2004 
February 2 and 3 using the VLA in its CnB configuration, except for the CB 65
observations that were carried out on 2005 February 12 in the BnA configuration.
 We selected the four IF spectral line mode to observe line 
and continuum simultaneously, processing both right and left circular 
polarizations, which we averaged together. Two IFs were used to observe the water maser transition, sampled on 64 channels over 
a bandwidth of 3.125 MHz, with a velocity resolution of 0.66 km~s$^{-1}$.  
The other two IFs were used for radio continuum observations at 1.3
cm, covering a 25 MHz bandwidth on 8 channels, and centered 50 MHz
above the central frequency used for line observations. 
The central velocity of the 
bands for line observations, the coordinates of the
 phase centers and the synthesized beam information are listed 
in Table~\ref{tbl-cal} for each source. The primary calibrators were 
3C48 (adopted flux density of 1.132 Jy) for observations 
on 2004 February 2, and 3C286 (adopted flux density 
of 2.539 Jy) for observations on 2004 February 3 and 2005 February 12.
The phase calibrators and their bootstrapped flux densities are given in 
Table~\ref{tbl-cal}. We used J0609-157, J1743-038, and 3C286 as
bandpass calibrators. Calibration and data reduction were performed with the 
Astronomical Image Processing System (AIPS) of NRAO. 
We detected water maser emission in CB 54, CB 101, and CB 232 (see
Table~\ref{tbl-masers}), but we
did not detect any emission towards either CB 65 or CB 205. 
In the case of CB 101 and CB 232, maser data were self-calibrated, and
spectral 
Hanning smoothing was applied (to alleviate ringing in the bandpass), 
which provides a final velocity resolution of $\sim$ 1.3 km s$^{-1}$. 
The continuum data of these two sources were cross-calibrated using the 
self-calibration solutions obtained for the stronger maser lines. 
None of the sources is detected in radio 
continuum  (see \S \ref{Res}), probably because the typical
values of the centimeter continuum emission observed in Bok globules 
(see \citealt{Mor97, Mor99}) fall below our sensitivity limit
 
We have also processed water maser data in source CB 3, taken from the VLA 
archive. The observations were carried out on 1995 October 28,
in the B configuration. These observations were
made in the 1IF spectral line mode, in right circular polarization only, 
with a bandwidth of 6.25 MHz sampled over 128 channels (velocity
resolution of 0.66 km~s$^{-1}$). The velocity of the 
center of the bandwidth, coordinates of the phase center, and
synthesized beam size are also listed in Table~\ref{tbl-cal}. The source 3C48 was used as primary 
flux calibrator, with an assumed flux density of 1.131 Jy, while
J0136+478 was used as phase and bandpass calibrator (Table~\ref{tbl-cal}).
Water maser emission was detected toward CB 3 (see Table~\ref{tbl-masers}).
The data were self-calibrated and spectral Hanning smoothing was applied, with a final velocity resolution 
of $\sim$ 1.3 km s$^{-1}$.  

\subsection{Single-dish observations}
\subsubsection{IRAM 30 m}
Millimeter single-dish observations were carried out towards CB 65
(IRAS 16277$-$2332) and CB 101 (IRAS 17503$-$0833) with the IRAM 30-m
telescope at Pico Veleta (Spain), in 2004 July-August. We have used 
Superconductor-Insulator-Superconductor (SIS) heterodyne
 receivers to observe nine different transitions 
at $\sim 1.3$, $\sim 2.7$,  and $\sim 3$ mm. 
We observed the $^{13}$CO($J=1\rightarrow0$), $^{13}$CO($J=2\rightarrow1$), C$^{18}$O($J=1\rightarrow0$), 
 C$^{18}$O($J=2\rightarrow1$), CO($J=1\rightarrow0$), CO($J=2\rightarrow1$), SiO($J=2\rightarrow1$), CS($J=2\rightarrow1$), and CS($J=5\rightarrow4$)
lines towards CB 101. In the case of CB 65, only the CO($J=1\rightarrow0$) and CO($J=2\rightarrow1$) lines were observed.
In Table \ref{tbl-cb101} 
we have summarized the rest frequencies of the different molecular 
transitions observed, the typical system temperature (T$_{\rm sys}$),
the half power beam width, the main beam efficiency, averaging area in
case the spectrum corresponds to the average of several positions, and
the derived line parameters. 
Pointing was checked every hour by observing J1743$-$038, giving a
pointing accuracy better that $2''$.
The observations were made by wobbling the secondary mirror to a
distance of 220$''$ from the source for CO($J=1\rightarrow0$), CO($J=2\rightarrow1$), and
CS($J=5\rightarrow4$) transitions, and in frequency switching mode for C$^{18}$O($J=1\rightarrow0$), 
C$^{18}$O($J=2\rightarrow1$), $^{13}$CO($J=1\rightarrow0$), $^{13}$CO($J=2\rightarrow1$), CS($J=2\rightarrow1$), and SiO($J=2\rightarrow1$) 
transitions. In addition, we observed CO($J=1\rightarrow0$) and CO($J=2\rightarrow1$) in frequency
switching mode at one selected position to better estimate the 
excitation conditions.
The data were taken with the versatile spectrometer assembly (VESPA) 
autocorrelator, split into two or three parts (depending on the lines), to
observe simultaneously two or three different frequencies. This 
provided resolutions between 0.05 and 0.4 km s$^{-1}$
at 1.3 mm, between 0.05 and 0.8 km s$^{-1}$ at 2.7 mm, and $\simeq 0.06$ km~s$^{-1}$
at 3~mm. 
Moreover, we used a 1 MHz filter bank split into two parts of
256 channels each, in combination  with VESPA. 
It provided a velocity resolution of 1.3 and 2.6 km s$^{-1}$
at 1.3 and 2.7 mm, respectively. 
The calibration was made using the
chopper wheel technique and the line intensities are reported as main
beam brightness temperatures.
With this setup, in some cases the same transition was observed with
different velocity resolutions. The values shown in Table
\ref{tbl-cb101} correspond to the data with best rms.

\subsubsection{Robledo de Chavela 70 m}
Centimeter single-dish spectral line observations of CB 101 were obtained with
NASA's 70 m antenna (DSS-63) at Robledo de Chavela, Spain, for both CCS
$J_{N}=2_{1}\rightarrow1_{0}$ and NH$_3$(1,1) transitions.
Rest frequencies, typical system temperatures, half power beam
widths, main beam efficiencies, and averaging region are given in 
Table ~\ref{tbl-cb101}. 
The rms pointing accuracy of the telescope was better than $6''$ and
$11''$ for CCS and ammonia observations, respectively.
The data were taken with a 1.3 cm receiver comprising a cooled 
high-electron-mobility transistor (HEMT) amplifier. A noise diode was used 
to calibrate the data. 
Observations were made in frequency switching mode, using a 256-channel
digital Fast Fourier Transform spectrometer. 
The CCS observations were performed on 2002 May
with a bandwidth of 1 MHz (velocity resolution $\simeq 0.05$ km s$^{-1}$),
while the NH$_3$ observations were carried out during 2003 July with
a bandwidth of 10 MHz (velocity resolution $\simeq 0.5$ km s$^{-1}$). 

All the single-dish data reduction was carried out using the CLASS package,
developed at IRAM and the Observatoire de Grenoble as part of the GAG
software.




\section{Results and discussion}
\label{Res}
\subsection{CB 3}
\begin{figure}
\rotatebox{0}{
\epsscale{0.95}
\plotone{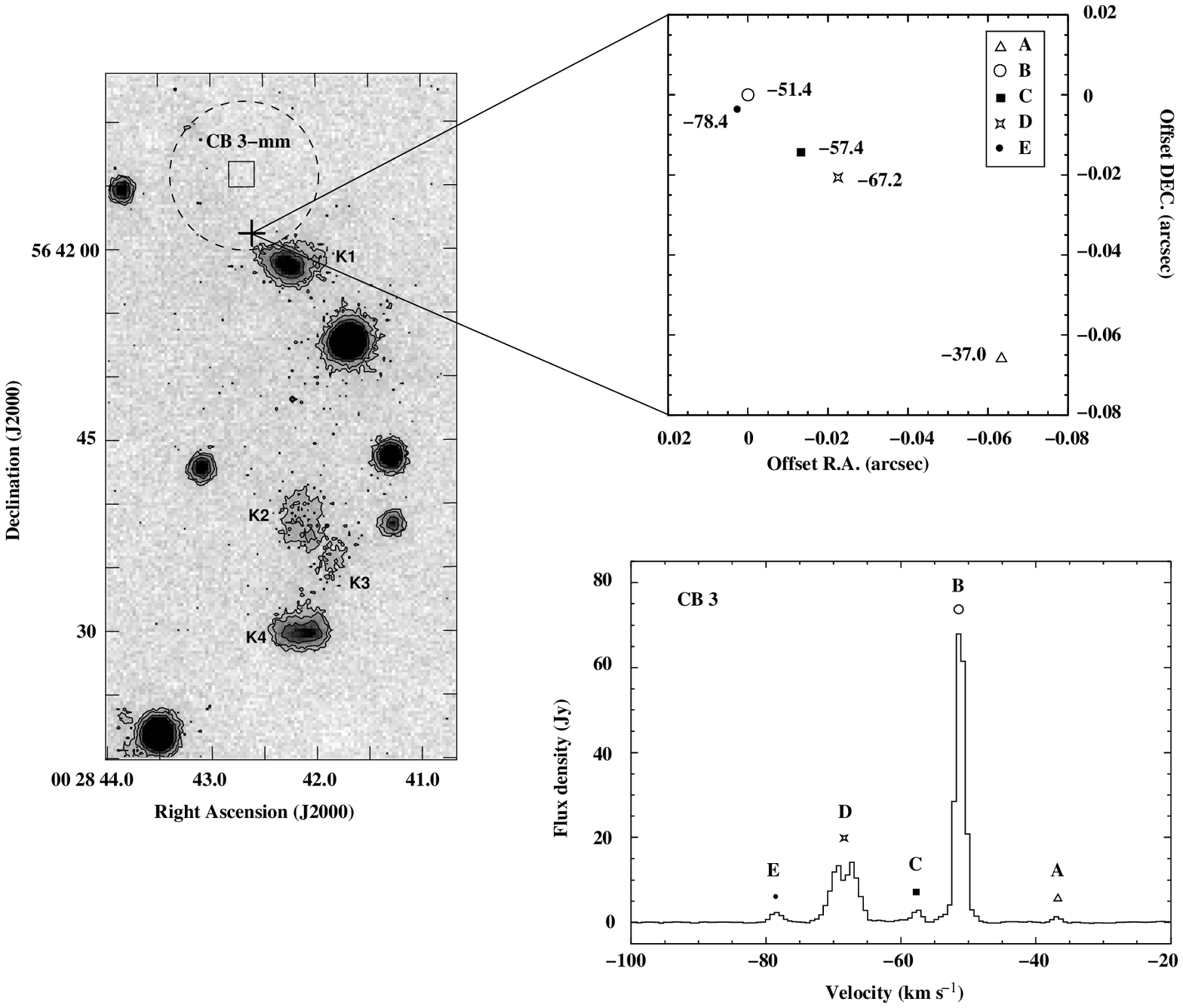}}
\caption{(Left) H$_{2}$ image (grey scale; \citealt{Mas04};
  astrometric accuracy of the image
  $\simeq1''$) of CB 3. K1, K2, K3, and K4 represent the H$_{2}$ knots
  detected by \citealt{Mas04}. The square marks the position of CB 3-mm
  \citep{Lau97} and the dashed circle represents the beamsize of their
  millimeter observations ($\simeq$12$\arcsec$). The cross marks the
  centroid position of the water maser emission detected with the
  VLA. (Lower right) Water maser spectrum of CB 3 obtained with the
  VLA. Different symbols correspond to the spectral features A, B, C,
  D, and E, indicated in Table~\ref{tbl-masers}. (Upper right) Spatial
  distribution of the independent maser spectral features showing a
  linear structure. The LSR
  velocity (in km s$^{-1}$) of each feature is shown. The (0,0) position in this map is the position of the
reference feature used for self-calibration
(Table~\ref{tbl-masers}).}
\label{cb3h2maser} 
\end{figure}

CB 3 is located at the near side of the Perseus arm, at a distance of
$\simeq$ 2.5 kpc \citep{Lau97,Wan95}. This globule shows a highly luminous FIR/submillimeter dust
condensation ($L_{bol}$ = 930 $L_{\odot}$), and it seems to be associated with intermediate-mass star formation \citep{Lau97,Lau97b}. Several sources in different stages of evolution have
been identified in CB 3: the young stellar object \objectname{CB 3/YC1}, which corresponds to \objectname{IRAS 00259+5625} \citep{Yun94b}, a near-infrared source, \objectname{CB 3 YC1-I}, 
that was proposed to be a Class II source  \citep{Yun95}, and a millimeter 
source, \objectname{CB 3-mm} \citep{Lau97}, cataloged by \citet{Cod99} as a probable 
Class 0 object, which also shows submillimeter emission \citep{Lau97b}.
This Bok globule is associated with a bipolar molecular outflow elongated in the
northeast-southwest direction \citep{Yun94b,Cod99}, of which CB 3-mm was proposed to
be the driving source \citep{Cod99}. The CO channel maps reveal 
different clumps along its main axis, which suggest episodic mass
loss \citep{Cod99}. There are four H$_{2}$ emission knots, projected towards
the blueshifted lobe of the outflow, and whose distribution does not
follow a straight line (see Fig.~\ref{cb3h2maser}), probably due to precession 
of the outflow axis \citep{Mas04}.

Our VLA maser spectrum is shown in Fig.~\ref{cb3h2maser}. Water maser emission in this source was first detected by
\citet{Scap91}, although the position reported with those single-dish
observations is shifted $\simeq$ (30$''$, $-$60$''$) from our VLA position.
At least five independent spectral features are evident in the VLA maser spectrum
(Fig.~\ref{cb3h2maser}), which we  designated as A,
B, C, D, and E on Table~\ref{tbl-masers}, and are centered at 
 $\simeq -37.0$, $-$51.4, $-$57.4, $-$67.2, and $-$78.4 km
s$^{-1}$ respectively. Component D may in its turn be composed of two
individual features, but they are blended together, given our spectral
resolution. All
except component A are blueshifted with respect to the cloud 
velocity ($V_{LSR}$=$-$38.3 km s$^{-1}$, \citealt{Cle88}). 
The five maser features delineate a spatial structure of 
$\simeq$ 0$\farcs$1 elongated from  northeast to southwest  
(see Fig.~\ref{cb3h2maser}). Assuming that the emission from a given
velocity channel is dominated by a single component, its position can
be determining with high accuracy. Therefore, the linear structure
observed is real and it is likely to be tracing the base of a jet, since the
relative positional uncertainty ($<$ 0$\farcs$007, see fourth column in Table~\ref{tbl-masers}) of the emission for individual maser spectral features is better than the total size of the structure ($\simeq$~0$\farcs$1; see Fig.~\ref{cb3h2maser}). 
The emission corresponding to the velocity closest to that of the cloud
(feature A) occupies the southern part of the structure, while that of the most
blueshifted one (E) is located at the north.

The maser emission is aligned in the same direction than the blueshifted molecular outflow,
and it is located
$\simeq$ 5'' south of the position of CB3-mm reported by
\citet{Lau97}, between 
this source and the northernmost H$_{2}$ knot observed by
\citet{Mas04} (see Fig.~\ref{cb3h2maser}). However, the mm observations
were made with a beamsize of $\simeq 12''$, so it is possible that
both mm and water maser emission actually come from the same location,
which would be more accurately traced by the masers. 
The proximity of the masers to the mm source is consistent with the
idea that the YSO traced by this source is exciting the water masers and probably powering the
molecular outflow. Nevertheless, although this mm source is the best candidate, we cannot rule out the existence of
another embedded source closer to the masers, and higher resolution
observations at millimeter and submillimeter wavelengths are needed in
order to clarify this point.

\begin{figure}
\epsscale{0.80}
\plotone{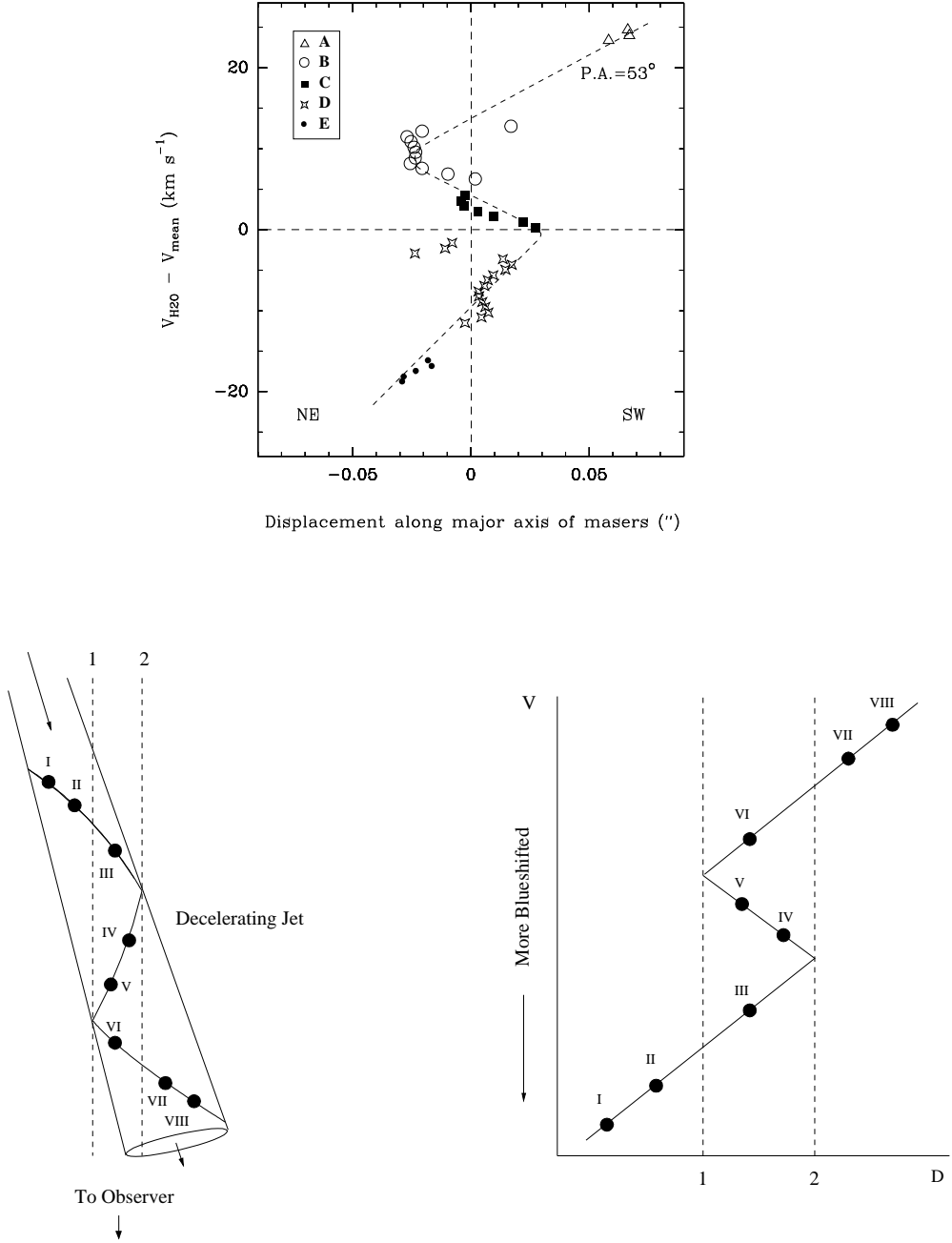}
\caption
{\scriptsize
{(Top) Position-velocity distribution of the centroids of the water maser
  emission at different velocity channels in 
  CB 3, along the major axis of the elongated maser
  distribution shown in Fig.~\ref{cb3h2maser}. The horizontal axis represents the relative  position of the emission along the major axis 
  (P.A. $=$ 53$\degr$) of the distribution with respect
  to their geometrical center. The vertical
  axis represents the velocity of each channel after
  subtracting the mean velocity of the spectrum ($-61.0$ km
  s$^{-1}$). The mean velocity is the sum of the velocity of the
  channels where maser emission is observed, divided by the total
  number of those channels. The geometric center (Offset R.A., Offset Dec.)=(0,0) is defined as the sum
  of the positional offsets (in R.A. or in Dec.) of the centroid of a single
  dominant component of water masers at each channel where emission is
  observed, divided by the total number of those channels. Different symbols correspond to  
the same spectral features indicated in Fig.~\ref{cb3h2maser}. The
dashed line 
represent the general trend shown by the maser spots in this
diagram. We note that points further away from the dashed
line are those with the largest positional uncertainty. 
(Bottom left) Model of a decelerating and 
  precessing jet. Filled circles
  represent hypothetical parcels of gas moving in the direction of the
  arrow. Dashed lines represent two lines of sight through the
  jet. The observer would be looking from the bottom of the page. (Bottom
  right)  Position-velocity distribution for the proposed model. Filled
  circles and dashed lines correspond to their equivalent in the left panel
  masers along the projected main axis of the jet. The filled circles do not try to reproduce the real water maser emission at different velocity channels, but the general tendency described by them (dotted line in top panel).}}
\label{cb3axis-model}

\end{figure}

In Fig.~\ref{cb3axis-model} we have represented the position-velocity
diagram of the maser emission along the major axis of the maser
structure, by considering the centroids of the maser emission for each
velocity channel. We note that although these centroids are not independent
spectral features, spatio-kinematical information of the
gas traced by water masers can be derived from the analysis of these
centroids, under the assumption that each velocity channel there is a
single dominant component. This diagram shows an interesting wave-like
distribution, similar to that observed in AFGL 2591 \citep{Tri03}.
\citet{Mas04}, based on the relative spatial distribution of H$_{2}$ knots with respect to the 
powering source, suggested the possible presence
of a precessing outflow.  Our position-velocity diagram also supports
this suggestion, but at smaller scales, with the wave-like distribution pattern compatible with a precessing jet traced by masers at scales of $\simeq$~250 AU. To illustrate this possibility we have represented an ideal model of a precessing jet that can qualitatively explain the same wave-like tendency observed in the position-velocity diagram. In the lower panels of Fig.~\ref{cb3axis-model}, 
 we show a sketch reproducing the model we propose. 
In order to fit the model with the observed pattern, we approximated
the precessing jet with a narrow cone on whose surface discrete
ejections of material are located and we assumed that the ejected
material is being decelerated (bottom left panel of
Fig.~\ref{cb3axis-model}). We also consider that the motion of the
ejected material is dominated by deceleration, while the changes in
velocity due to precession are negligible. Therefore, our
 model can reproduce a wave-like distribution
(solid lines joining the dots in the bottom right panel of
Fig.~\ref{cb3axis-model}) similar to the one observed (dashed lines
joining the symbols in the upper panel of Fig.~\ref{cb3axis-model}).
Under these assumptions, we estimate a deceleration of the ejected
material of 
$\simeq$~2.2~$\times$~10$^{-8}$ cos~$i$ km~s$^{-2}$ ($\simeq$~0.14 cos~$i$
AU~yr$^{-2}$), where $i$ is the position angle between the direction of
the jet and the plane of the sky.
 We caution that our aim with this model is to explain the wave-like tendency observed with the position-velocity diagram, and not to reproduce exactly the observed maser emission.
Further support for the possible presence of a precessing jet at these small
scales (250 AU) could be obtained with proper motions studies of masers by means of VLBI 
observations.

\subsection{CB 54}

\begin{figure}
\epsscale{1}
\plotone{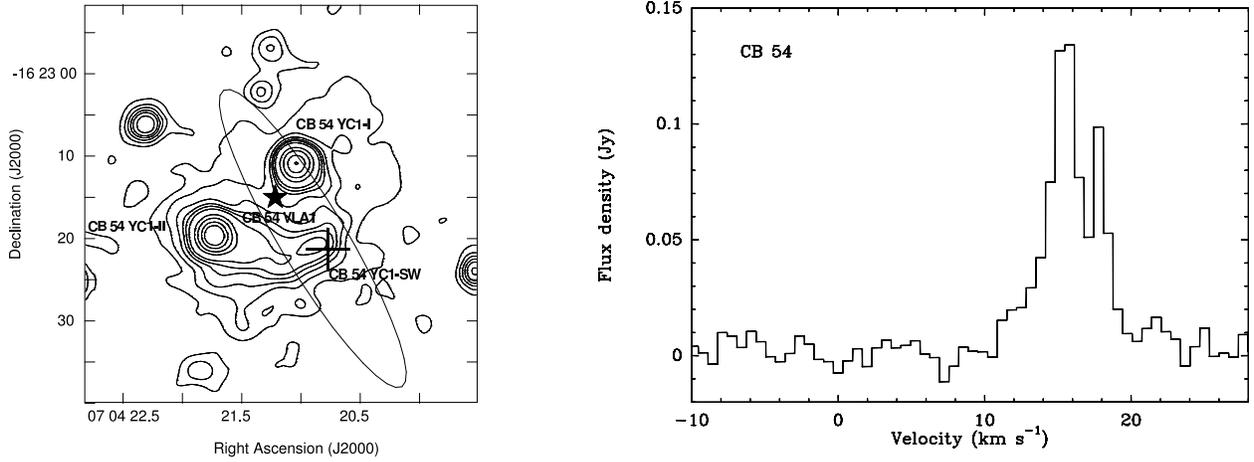}
\caption{(Left panel) Contour map of the 2MASS K-band emission (positional error
  $\simeq0$\farcs$2$) surrounding IRAS 07020$-$1618 in CB 54. The ellipse is
  the position error of the IRAS source. The cross indicates the
  centroid position of the water masers observed
  with the VLA. The star is the position of CB 54 VLA1, the radio continuum source at 3.6
  cm reported by Yun et al. (1996; positional error
  $\simeq2''$). (Right panel) Water maser spectrum of CB 54 obtained with the VLA.}
  \label{cb54composition} 
\end{figure}

This Bok globule hosts a multiple system of YSOs towards
\objectname{CB54 YC1} (\objectname{IRAS 07020-1618}), with the presence
of two bright near-infrared (K band, 2.2~$\mu$m) objects
(\objectname{CB54 YC1-I}, \objectname{YC1-II}, which are probably Class I protostars), plus a
bright elongated feature (hereafter \objectname{CB54 YC1-SW}) mainly seen in H$_{2}$ 
[v = 1--0 S(1), 2.121 $\mu$m] (\citealt{Yun94}, \citeyear{Yun95}; \citealt{Yun96}; see Fig.~\ref{cb54composition}).
In addition, \citet{Yun96} and \citet{Mor97} reported a radio continuum source
(CB 54 VLA1) at 3.6 and 6 cm located $\simeq$ 5$''$ to the NE of the
nominal position of the IRAS source
(see Fig.~\ref{cb54composition}). These sources are located near the center of a bipolar
CO outflow that is oriented in the northeast-southwest direction and probably
moves close to the plane of the sky \citep{Yun94b}.

Our VLA observations (Fig.~\ref{cb54composition}) reveal water maser emission with two
distinct spectral
features, at $\simeq 15.8$  and 17.8 km s$^{-1}$.
These velocities are
blueshifted with respect to the velocity of the cloud ($V_{LSR}$=19.5 km
s$^{-1}$; \citealt{Cle88}). 
No maser emission was detected  with the
Robledo antenna at the velocities reported here
\citep{Gom06}, which is understandable since the flux density
reported here is below the sensitivity threshold of those single-dish
observations. However, \citet{Gom06} detected a
component
at $\simeq 8$ km s$^{-1}$, which reached a flux density of up to
$\simeq 45$ Jy,
and another one of up to $\simeq 1$ Jy at $\simeq 14$ km s$^{-1}$. Neither of these is
evident in the VLA spectrum (Fig.~\ref{cb54composition}), although
the component
at $\simeq 14$ km s$^{-1}$ might be present at a very low level, and blended with
the one at $\simeq 15.8$ km s$^{-1}$.

The maser emission is located at the position of
the   elongated feature  CB54 YC1-SW
(see Fig.~\ref{cb54composition}). \citet{Yun96b}, on the basis that the
elongated feature appears brighter in H$_2$ than in the K-band, proposed
that it could trace shocked material, such as a knot in a
near-infrared jet. However, given the association of CB 54 YC1-SW with
the water masers, we suggest that this feature is another embedded YSO
and that, according to the association of water masers with mass-loss
phenomena,  this
object would be a good candidate for being the powering source of the observed 
molecular outflow in the region. 
In fact, masers are located $\simeq$ 18000-20000 AU (at a distance of 1.5~kpc; \citealt{Lau97}) from both 
CB 54 YC1-I and  CB 54 YC1-II, which make these objects less
likely candidates for pumping the maser emission, since masers in
low-mass star-forming
regions tend to be within several hundred AU from the powering source
\citep{Che95, Cla98, Fur00, Fur03}. We did not detect 1.3 cm continuum emission
with the VLA either at the position of CB54 YC1-SW or toward the other
proposed YSOs in the region, with a 3$\sigma$ upper limit of 0.4 mJy. Deeper
radio continuum measurements in this region could help to confirm whether
CB54 YC1-SW is indeed a YSO.

\subsection{CB 65}
\objectname{IRAS 16277$-$2332} is located at the north-western edge of the Bok 
globule CB 65, in Ophiucus. The nature of this IRAS source is  unknown, and 
in fact, \citet{Par88} suggested that it may not be associated with the
globule. \citet{Vis02} carried out submillimeter observations towards
CB 65, and detected a submillimeter core at the center of the globule at
a distance of $\simeq 3'$ southeast of IRAS 16277$-$2332, but no
emission was detected at the IRAS position.
Our single dish  survey with the Robledo de Chavela antenna
\citep{Gom06} revealed water maser emission in IRAS
 16277$-$2332, near the CB 65 cloud velocity (V$_{LSR}$ = 2.3 km
 s$^{-1}$; \citealt{Cle88}), with a peak flux density of 0.3 Jy on
 2002 June 16. 
However, no water maser emission was detected with the VLA ($\leq$ 40
mJy, 3$\sigma$ upper limit), which is not surprising given the time variability of these masers \citep{Rei81}. 
No 1.3 cm continuum emission was detected, with a 3$\sigma$ upper limit of 2.7 mJy.   
On the other hand, the observations performed with the IRAM 30~m antenna in the CO(1$-$0)
and (2$-$1) transitions, showed no high-velocity wings in the
spectra, with an rms of 0.03~K and 0.3~K  
respectively. This indicates the absence of any significant mass-loss
activity.  

Unfortunately, there is not enough information in the literature about IRAS 16277$-$2332
that may help us to clarify the real nature of this source. For
instance, IRAS data show emission at 60~$\mu$m, but only  upper limits
at 12, 25, and 100~$\mu$m.
 There are no Midcourse Space Experiment (MSX) infrared data
 nor emission in the 2MASS K, H, or
J bands, from which to obtain information about its spectral energy 
distribution. Deeper infrared observations are needed to reveal the
nature of this source.

\subsection{CB 101}
CB 101 is a Bok globule located at 200 pc \citep{LeeM99}, and cataloged as a cold and quiescent cloud 
\citep{Cle91} that hosts two IRAS sources, \objectname{IRAS 17503$-$0833} and 
\objectname{IRAS 17505$-$0828}. 
Our water maser single-dish survey \citep{Gom06} revealed water maser emission towards 
IRAS 17503$-$0833. This source is located $\simeq$ 9$'$ south of the globule center. 

\subsubsection{Water Masers and Radio Continuum Emission}
\begin{figure}
\epsscale{1}
\plotone{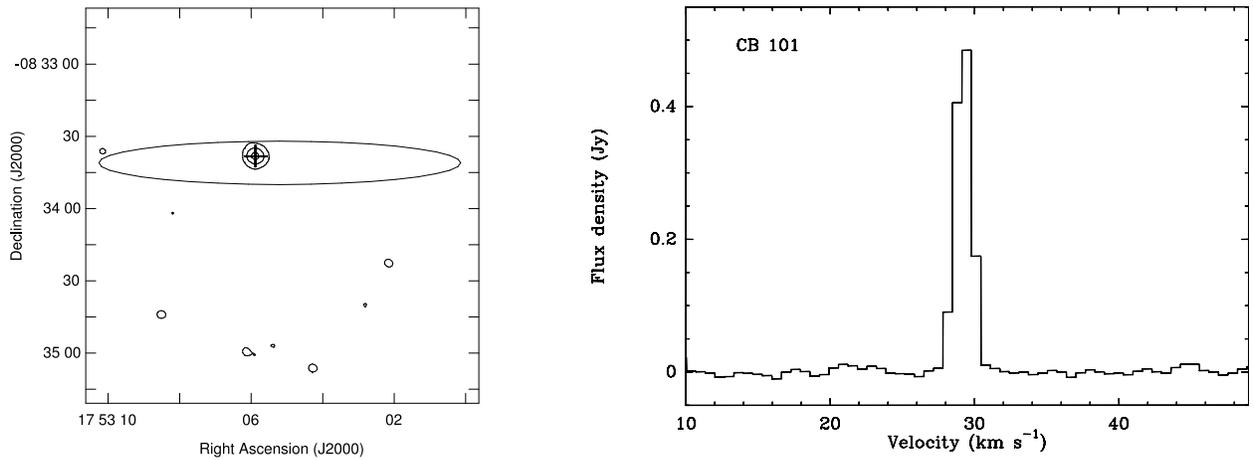}
\caption{(Left panel) Contour map of the K-band 2MASS emission (positional error 
  $\simeq0$\farcs$14$) around IRAS 17503-0833 in CB 101. The cross
  represents the centroid position of the water maser emission. The ellipse represents the position error of IRAS 17503-0833. (Right panel) Water maser spectrum toward IRAS 17503$-$0833 in CB101, obtained with the VLA.}\label{cb101maser-k}
\end{figure}

We have detected with the VLA water maser emission toward IRAS 17503$-$0833 
at V$_{LSR}$ $\sim$ 29.5 km s$^{-1}$ (see Table~\ref{tbl-masers} and
 Fig.~\ref{cb101maser-k}), which is 
redshifted with respect to the velocity of the molecular cloud
($V_{LSR}$ = 6.7 kms$^{-1}$; \citealt{Cle88}). The flux density and
velocity of the water maser emission are similar to those found with
the Robledo antenna \citep{Gom06}. The water maser
emission coincides with a point source observed in the 2MASS K-band, 
which is probably the near-IR counterpart of IRAS 17503$-$0833 
(see Fig.~\ref{cb101maser-k}).  
We did not detect VLA radio continuum emission 
at 1.3 cm, with a 3$\sigma$ upper limit of 0.4 mJy. 

\subsubsection{Millimeter and Centimeter Single-Dish Observations}
\begin{figure}
\epsscale{1}
\plotone{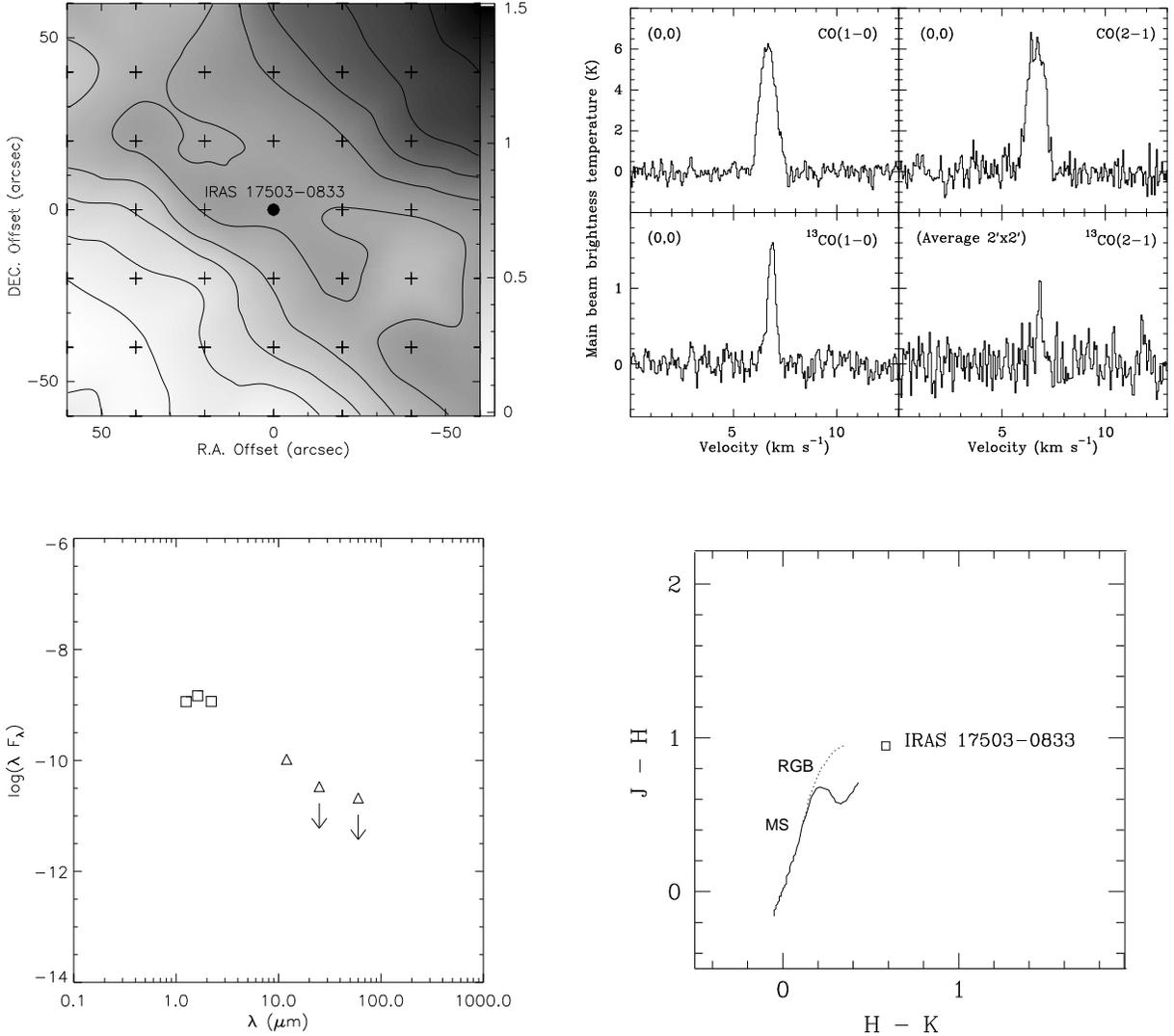}
\caption{(Upper left) Map of the $^{13}$CO($J=1\rightarrow0$) integrated emission in CB 101. The
  reference position (filled circle) is that of IRAS 17503-0833. Contour levels are
represented from 0.15 to 1.5 K km s$^{-1}$ in steps of 0.15 K km
s$^{-1}$. Crosses mark the observed positions. (Upper right) Spectra of CO($J=1\rightarrow0$), CO($J=2\rightarrow1$), $^{13}$CO($J=1\rightarrow0$), and $^{13}$CO($J=2\rightarrow1$) at the position of IRAS 17503-0833 in CB 101. The
  $^{13}$CO($J=2\rightarrow1$) spectrum (lower right) is the average over an area of
  $\simeq 2' \times 2'$ centered at the position of the IRAS source.
(Lower left) Infrared spectral energy distribution of IRAS 17503-0833 in
  CB 101. The squares represent the 2MASS measurements in the H, J,
  and K band and the triangles represent the IRAS measurements at 12,
  25, and 60 $\mu$m (see Table~\ref{tbl-cb101_ir}). (Lower right) 
  Location of IRAS 17503$-$0833 in an (H-K)-(J-H) color
  diagram. The square represents the dereddened infrared color of 
  IRAS 17503$-$0833. The solid line represents the main sequence, and the dotted line represents the Red Giant Branch.}
\label{cb101varios}
\end{figure}

Our aim was to search for
any sign of star formation activity toward IRAS 17503$-$0833 (e.g.,
presence of molecular outflows, high density molecular gas, and/or shocked
material). We mapped the CO($J=1\rightarrow0$) and ($J=2\rightarrow1$) transitions over an area of
$40''\times 40''$ in order to detect a possible molecular outflow in the
region. No high-velocity emission was 
detected in the spectra of any of these transitions, with a rms of
0.012 K per channel (channel width $\simeq$ 0.8 km~s$^{-1}$) and
0.026 K per channel (channel width $\simeq$ 0.4 km~s$^{-1}$), respectively. Moreover, as for the general distribution of the molecular gas,
Fig.~\ref{cb101varios} shows the map of $^{13}$CO($J=1\rightarrow0$) integrated
intensity in a region of $2'\times 2'$. It reveals more intense emission towards the northwest
of the IRAS source, which probably originates from the molecular gas contained
in the Bok globule, but there is no obvious local maximum of molecular
gas towards the source. 
In Table~\ref{tbl-cb101} we summarized the parameters of the single-dish spectra 
of different molecular transitions towards IRAS 17503$-$0833. 

The $^{13}$CO($J=2\rightarrow1$) emission is weak, and barely detectable in the region
(Fig.~\ref{cb101varios}).
We did not detect emission of high-density molecular gas
tracers such as the CCS(2$_{1}$-1$_{0}$), NH$_{3}$(1,1), CS($J=2\rightarrow1$) and
CS($J=5\rightarrow4$) lines. No emission of the  SiO($J=2\rightarrow1$) transition was detected
either. This transition is usually associated 
with shocked regions around young stellar objects \citep{Har98,Gib04}.

We derived the physical conditions from the CO and $^{13}$CO spectra
obtained towards the IRAS source, under the following three assumptions \citep{Est96}: (1) local thermodynamic
equilibrium, (2) the CO emission is optically thick, and  (3) $^{13}$CO emission is optically thin.
The CO($J=1\rightarrow0$) emission towards the IRAS position was used to obtain the
excitation temperature ($T_{\rm ex}$) from:

$$T_{\rm ex} = 5.53 \left[\ln \left( 1 + \frac{5.53}{T_{\rm CO} + 0.82}\right)\right]^{-1}\; K,$$ 

\noindent
where $T_{\rm CO}$ is the main beam brightness temperature of the CO($J=1\rightarrow0$) emission, in K.

 For the $^{13}$CO($J=1\rightarrow0$) we derived its optical depth ($\rm\tau_{\rm
  ^{13}CO}$) and its column density $N$($^{13}$CO) as:

$$\tau_{\rm ^{13}CO} = -\ln \left[1-T_{\rm ^{13}CO}
  \left(\frac{5.29}{\exp \left(5.29/T_{\rm ex}\right)-1}-0.87\right)^{-1}\right],$$

$$N({\rm ^{13}CO})=2.42\times10^{14}\frac{T_{\rm ex} \Delta v ~\tau_{\rm ^{13}CO}}{1-\exp\left(-5.29/T_{\rm ex}\right)}\; {\rm cm^{-2}},$$

\noindent
where $T_{\rm ^{13}CO}$ is the main beam brightness temperature of the $^{13}$CO($J=1\rightarrow0$) emission in K, and 
$\Delta v$ is the line width at half maximum, in km s$^{-1}$. 

The physical parameters (derived from the spectra shown in
Fig.~\ref{cb101varios} and their corresponding values in Table~\ref{tbl-cb101}), are $T_{\rm ex}=9.6\pm0.3$ K,
$\tau_{\rm ^{13}CO}=0.27\pm 0.07$, and $N(\rm ^{13}CO)=(6.6\pm 2.6)\times 10 ^{14}$ cm$^{-2}$ at the position of the IRAS
source. The value of $N(\rm H_{2})$ is (3.3$\pm$1.2)$\times$10$^{20}$
cm$^{-2}$, derived from the relative abundance of $^{13}$CO with
respect to H$_{2}$, proposed by \citet{Dic78}, [H$_{2}$/$^{13}$CO]=5$\times$10$^{5}$.
\citet{Kim02} reported a value of N($\rm^{13}CO$)=3.7$\times$10$^{15}$ cm
$^{-2}$ towards the central peak position of CB 101 (located
$\simeq9\arcmin$ north). This value at the center of the globule is $\simeq$~6
times higher than the value derived by us at the position of IRAS
17503$-$0833, which is consistent with a decrease of the gas density towards
the edges of the cloud.

\subsubsection{The Nature of IRAS 17503$-$0833} 
The absence of any signpost of star formation activity associated with the 
water maser emission, suggests that the water maser in CB 101 may
not be associated with a young object but with an evolved star, since
water maser emission is also known to be associated with circumstellar
envelopes of late-type stars \citep{Bow84, Eng86, Eng88, Hab96}. Moreover, the
difference between the centroid  velocity of the detected water maser and the
cloud velocity is $\simeq 23$ km s$^{-1}$, whereas this difference is usually
$\leq 15$ km s$^{-1}$ in the case of YSOs \citep*{wil94,Ang96,bra03}. 
To further investigate the true nature of IRAS 17503$-$0833, we
searched for additional data in the literature.


The IRAS flux density at 12~$\mu$m is well determined, but the values for 25, 60, and
100~$\mu$m are only upper limits. The near-infrared flux densities were retrieved
from the 2MASS catalog, and dereddened with a value for E(B$-$V) = 0.21, 
taken from \citet{Whi86} for stars with galactic latitude
b = 8~$\deg$. The corrected J, H, and K magnitudes are: 8.643, 7.690, and 7.100,
respectively. The near- and far-infrared data flux densities 
are shown
in Table~\ref{tbl-cb101_ir}, where we do not include the IRAS flux
density
at 100~$\mu$m due to the low quality of those data.

In Fig.~\ref{cb101varios} we have plotted the spectral energy
distribution (SED) of IRAS 17503$-$0833, including the upper 
limits for the IRAS flux densities. 
Evolved objects may show one or two maxima in their SED. After the AGB the 
SEDs usually show two maxima, one belonging to the central 
star and the other to the circumstellar envelope \citep{Kwo93}.
In the case of IRAS 17503$-$0833, if it is an evolved object 
there seem to be only one maximum in the near infrared; the second maximum should have shown up in the far infrared. 
The presence of only one maximum suggests that no disk
or circumstellar envelope is detached from the central star. This
would restrict the evolutionary stage of the star to the phase between Red
Giant and the Asymptotic Giant Branch (AGB).
The SED is then compatible with a black body distribution at a temperature of $\simeq 2200$ K, although more data points at different wavelengths would be necessary to better constrain this temperature.
In Fig.~\ref{cb101varios}, we have plotted the dereddened IR colours 
of the object in an IR color-color diagram, to further check its evolutionary
stage. IRAS 17503$-$0833 is located to the right and above the Main 
Sequence and the Red Giant Branch, which is typical of Mira stars (see
\citealt{Whi94, Whi95}). The temperatures of these stars fall in the range between 2000~K and 
3500~K \citep{vanB02} , which is compatible with the temperature of
$\sim 2200$~K estimated by us. Moreover, 
water maser emission is very frequent in Mira-type stars \citep{Bow84}. 
Therefore, we suggest that the CO emission found belongs to the
cloud CB 101, and IRAS 17503$-$0833 is a field star, not associated
with the cloud. This is consistent with the location of this source at
the edge of the Bok globule, at a distance of $\simeq 9\arcmin$
from its center. 
In any case, additional observations at other frequencies, in order to be able to accurately determine the effective temperature of the star from the SED,
together with a light curve to perform  variability studies of this
source are needed to confirm its identification as a Mira star.

\subsection{CB 205}

\objectname{CB 205}, at a distance of 2 kpc \citep{Lau97}, is a very active star formation region containing several YSOs
\citep{Her76,Nec85,Nec90,Hua00,Mas04}, and a weakly collimated bipolar
molecular outflow with a significant overlapping of the redshifted and
blueshifted emission lobes \citep{Xie90}.

CB 205 is one of the few Bok globules in the CB catalog where water maser
emission was known \citep{Nec85} before the survey of \citet{Gom06}. 
The maser, located near \objectname{IRAS 19433+2743}, showed
a peak flux density of 
$\simeq$ 4.8 Jy on May-October 1983. This 
emission was detected again by \citet{Bra94} on 1992 January 18, with a
peak flux density of 0.7 Jy and located $\simeq$ 40$''$ west of 
our phase center, although inside the 2$'$ VLA primary beam at 22 GHz.
However, \citet{Gom06} did not detect the maser, with a 
$3\sigma$ upper limit of 0.75 Jy on 2003 July 11 and of 
0.15 Jy on 2004 July 22 and 2005 June 16. 
 Our VLA observations also failed to detect any water maser emission
 on 2004 February 3, at a level of 
$\ga 7$~mJy (3$\sigma$).
In our VLA observations, we did not detect any radio continuum
emission at 1.3 cm, with a 3$\sigma$  
upper limit of 0.5 mJy.

\subsection{CB 232}
This Bok globule contains a CO bipolar molecular outflow centered near
\objectname{CB 232 YC1} (\objectname{IRAS 21352+4307}), 
whose lobes are slightly overlapped and exhibit a
poor degree of collimation \citep{Yun94b}.
Near-infrared maps reveal a single source, the possible
counterpart of IRAS 21352+4307 \citep{Yun94},  which was designated as
\objectname{CB 232 YC1-I} and classified as a Class I 
object by \citet{Yun95}. 
\citet{Hua99} detected two submillimeter sources, SMM1 (the westernmost 
one) and SMM2 (the easternmost one), located $\simeq$ 10$''$ west
and $\simeq$ 5$''$ south-east from CB 232 YC1-I respectively, with a positional 
error of $\simeq4''$. SMM1 was classified as a Class 0 source while SMM2 was proposed to be either 
the submillimeter counterpart of CB 232 YC1-I, or another Class 0 source without
infrared counterpart \citep{Hua99}.

\begin{figure}
\epsscale{1}
\plotone{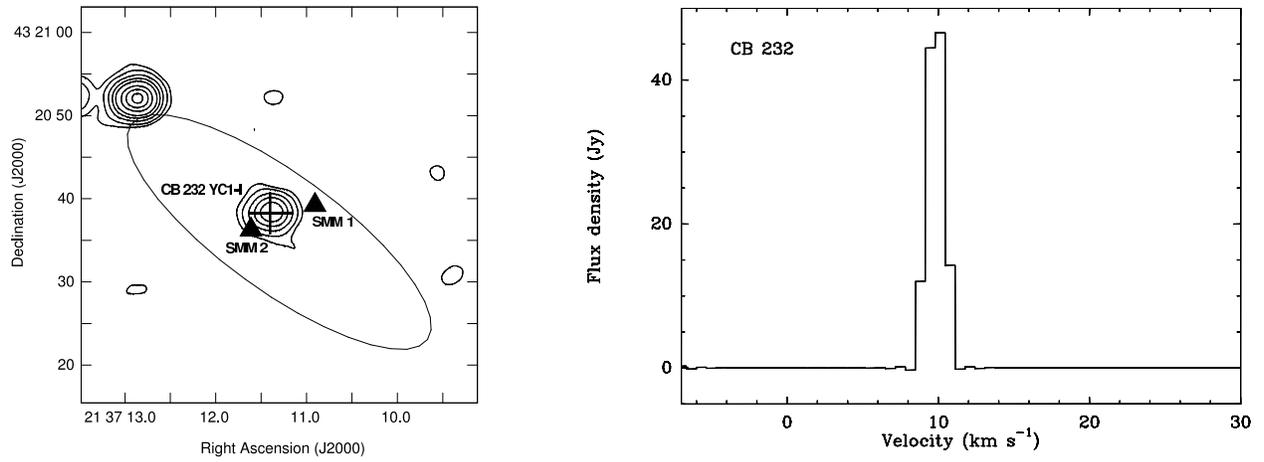}
\caption{(Left panel) Contour map of the near-infrared
  emission in K-band 
  (astrometric accuracy of the image $\simeq0$\farcs$12$) around IRAS 21352+4307. The
  ellipse represents the error in the position of the IRAS source. 
  The cross represents the position of the water maser emission, and
  the triangles mark the position of the submillimeter sources reported by
  Huard et al. (1999; positional error $\simeq4''$). (Right panel) Water
  maser spectrum of CB 232, obtained with the VLA. }\label{cb232maser-k}
\end{figure}

Our VLA observations show water maser emission with velocity V$_{LSR}$
$\sim$ 10.1 km s$^{-1}$ (see Table~\ref{tbl-masers} and Fig. \ref{cb232maser-k}), close to the
cloud velocity ($V_{LSR}$=12.6 km s$^{-1}$; \citealt{Cle88}), and
coinciding, within the positional uncertainty, with the position of the  
near-infrared source CB 232 YC1-I (see Fig.~\ref{cb232maser-k}). 
This positional agreement favors CB 232 YC1-I as the most
likely candidate to power the maser emission and
the molecular outflow observed in the
region (according to the typical association of water masers to
mass-loss phenomena). 

The velocity of
the maser is consistent with that found in the single-dish
detection by \citet{Gom06}, although the flux density was a
factor of $\sim 50$ larger at the epoch of our
interferometric observations. 
We did not detect radio continuum emission at 1.3 cm, with a 3$\sigma$ upper limit 
of 0.5 mJy.

\section{Conclusions}
In this work we have presented interferometric observations using the
VLA,  of
several sources  of water 
maser emission detected as part of a single-dish survey 
performed with the Robledo-70m antenna towards Bok globules. 
The general conclusions we have obtained are the following:

\begin{itemize}
\item We have detected water maser emission with the VLA in CB 3, CB 54, 
 CB 101 and CB 232. No water maser emission was detected towards CB 65 or CB 205. 
 
\item These water masers  are associated with 
star forming regions showing 
 bipolar  
molecular outflows, except for CB 101 (IRAS 17503$-$0833), which we propose is
an evolved object, possibly a Mira star, and CB 65 (IRAS
16277$-$2332), whose nature is unknown.

\item All of the Bok globules associated with star formation present
  multiple  
  stellar systems in different evolutionary stages. Accurate positions
  of water  
  masers has helped us to get information about the powering sources
  in the case 
  of CB 3, CB 54, and CB 232. We propose the millimeter source CB 3-mm
and the near-IR source 
  CB 232 YC1-I
 as the best candidates for pumping the maser emission in CB 3 and CB 232 respectively.
 In the case of CB 54, we propose a new young 
  object (CB 54 YC1-SW), seen as an  elongated feature in
  near-infrared images, as the powering source of the maser emission.

\item The maser emission in CB 3 is distributed along the direction of
  the molecular outflow, and seems to be tracing the inner part of a
  collimated jet. The position-velocity distribution of the maser emission shows a
  wave-like structure, which is consistent with a precessing
  jet.

\end{itemize}

\acknowledgments

The authors thank the valuable comments of our anonymous referee, which have greatly improved  this work.
We are also thankful to Fabrizio Massi for providing the
near-infrared data in CB~3 and David Barrado y Navascu\'es for his help with the
data reduction of this image. 
GA, JFG, and JMT acknowledge support from the Ministry of
Science and Technology (MCYT) grant (European Fund of Regional
Development, FEDER) AYA2005-08523-C03 (Spain), and OS acknowledges support
from the grant AYA 2003-09499.
GA and JFG acknowledge support from Junta de Andaluc\'{\i}a (TIC-126).
IdG acknowledges the support from a Calvo Rod\'es
Fellowship from the Instituto Nacional de T\'ecnica Aeroespacial (INTA).
The work by TBHK was done at the Jet Propulsion Laboratory, California
Institute of Technology, under contract to the National
Aeronautics and Space Administration (NASA). JMT acknowledges the
high hospitality of the UK Astronomy Technology Centre, Royal
Observatory Edinburgh, during his sabbatical stay. 
This paper is partly based on observations taken during 
``host-country'' time allocated at Robledo de Chavela; this time is
managed by the Laboratorio de Astrof\'{\i}sica Espacial y F\'{\i}sica
Fundamental (LAEFF) of INTA, under agreement with National
Aeronautics and Space Administration/Ingenier\'{\i}a y Servicios
Aeroespaciales (NASA/INSA). It also makes use of data products from
the Two Micron All Sky Survey (2MASS), which is a joint project of the
University of Massachusetts and the Infrared Processing and Analysis
Center/California Institute of Technology, funded by NASA and the
National Science Fundation.



\clearpage





\clearpage

\begin{deluxetable}{lrrcrcccc}
\tabletypesize{\scriptsize}
\tablecaption{Setup of VLA observations\label{tbl-cal}}
\tablewidth{0pt}
\tablehead{
\colhead{Source}&
\colhead{R.A.\tablenotemark{a}}&
\colhead{Dec.\tablenotemark{a}}&
\colhead{$V_{0}$\tablenotemark{b}} &
\colhead{Beam Size} &
\colhead{Beam P.A.}   &
\colhead{Phase calibrator}& 
\colhead{$S_{\rm cal}$\tablenotemark{c}}& 
\colhead{Date\tablenotemark{d}}
\\
 & 
\colhead{(J2000)}&
\colhead{(J2000)}&
\colhead{(km s$^{-1}$)}&
\colhead{($\arcsec$)} &
\colhead{($\deg$)}&  
& 
\colhead{(Jy)} 
&
}
\startdata
CB 3\tablenotemark{e}   &00 28 43.5075  &$+$56 41 56.868  &$-$60.0   &0.55$\times$0.25  &81 &J0136$+$478   &5.8$\pm$1.2    &95/10/28   \\
CB 54                   &07 04 21.2170  &$-$16 23 15.000  &7.9       &0.93$\times$0.44  &73 &J0609$-$157   &4.55$\pm$0.17  &04/02/02   \\
CB 65                   &16 30 43.7109  &$-$23 39 07.736  &2.3       &0.28$\times$0.18  &86 &J1626$-$298   &2.00$\pm$0.03  &05/02/12   \\
CB 101                  &17 53 05.2300  &$-$08 33 41.170  &28.8      &0.99$\times$0.39  &68 &J1743$-$038   &7.04$\pm$0.05  &04/02/03   \\
CB 205                  &19 45 23.8630  &27 50 57.840     &13.2      &0.86$\times$0.79  &48 &J2015$+$371   &3.06$\pm$0.05  &04/02/03   \\
CB 232                  &21 37 11.3100  &43 20 36.260     &12.1      &1.08$\times$0.67  &53 &J2202$+$422   &2.50$\pm$0.04  &04/02/03   \\
\enddata
\tablenotetext{a}{Coordinates of the phase center. Units of right ascension are hours, 
minutes, and seconds. Units of declination are degrees, arcminutes, and arcseconds.}
\tablenotetext{b}{$\rm L$ocal standard of rest velocity of the center of the bandwidth.}
\tablenotetext{c}{Bootstrapped flux densities of phase calibrators at
  22 GHz. Uncertainties are 2$\sigma$.}
\tablenotetext{d}{Observation date, YY/MM/DD}
\tablenotetext{e}{Archive data}
\end{deluxetable}

\begin{deluxetable}{lccccr}
\tabletypesize{\scriptsize}
\tablecaption{Water maser features detected with the VLA. \label{tbl-masers}}
\tablewidth{0pt}
\tablehead{
\colhead{Source}&
\colhead{Offset R.A.\tablenotemark{a}} &
\colhead{Offset Dec.\tablenotemark{a}}& 
\colhead{Position\tablenotemark{b}} &
\colhead{Flux density\tablenotemark{c}}   &
\colhead{$V_{\rm LSR}$\tablenotemark{d}}
\\
 &  
\colhead{($\arcsec$)}&
\colhead{($\arcsec$)}& 
\colhead{uncertainty ($\arcsec$)} &\colhead{(Jy)}& 
\colhead{(km s$^{-1}$)} 
}
\startdata

CB 3 (A)        &$-$0.063    &$-$0.065        &0.007                    &1.29$\pm$0.03      &$-37.0$\\ 
\phm{CB 3} (B)  &0           &0               &\nodata\tablenotemark{e} &68.19$\pm$0.04        &$-51.4$\\
\phm{CB 3} (C)  &$-$0.013    &$-$0.014        &0.003                    &2.77$\pm$0.03      &$-57.4$\\
\phm{CB 3} (D)  &$-$0.0226   &$-$0.0207       &0.0006                   &19.78$\pm$0.04         &$-67.2$\\ 
\phm{CB 3} (E)  &0.003       &$-$0.003        &0.003                    &2.86$\pm$0.03      &$-78.4$\\ 
CB 54           &$-$6.49     &$-$6.30         &0.03                     &0.087$\pm$0.005  &$17.8$\\
                &$-$6.467    &$-$6.293        &0.018                    &0.129$\pm$0.005  &$15.8$\\
CB 101          &0   &0    &\nodata\tablenotemark{e} &0.499$\pm$0.003  &$29.5$\\
CB 232          &0   &0    &\nodata\tablenotemark{e} &46.568$\pm$0.006     &$10.1$\\
\enddata
\tablenotetext{a}{Position offsets of the peak of each distinct water maser feature with respect to the reference feature
used for self-calibration, which are (R.A., Dec.)$_{J2000.0}$ = (00 28
42.612, 56 42 01.17) for CB 3, (17 53 05.882, $-$08 33 38.17) 
for CB 101 and (21 37 11.402, 43 20 38.26 ) for CB 232, 
or with respect to the phase center (07 04 21.217, $-$16 23 15.00) for
CB 54. Units of right ascension are hours, 
minutes, and seconds. Units of declination are degrees, arcminutes, and arcseconds.}
\tablenotetext{b}{Uncertainty in the relative positions with respect to the
  reference positions. The absolute
  position uncertainty of the reference position is $\simeq$ 0$\farcs$19 for CB 3,
  $\simeq$ 0$\farcs$05 for CB 101, and $\simeq$ 0$\farcs$12 for CB 232 and CB 54. Uncertainties are 2$\sigma$.}
\tablenotetext{c}{Quoted uncertainties are two times the rms noise in the
  maps.}
\tablenotetext{d}{\rm LSR velocity of the spectral features. Velocity resolution $\sim$1.3 km~s$^{-1}$.}
\tablenotetext{e}{Reference feature.}
\end{deluxetable}

\begin{deluxetable}{lcrccccccccc}
\tabletypesize{\scriptsize}
\rotate
\tablecaption{Single-dish observations towards CB 101 and CB 65 \label{tbl-cb101}}
\tablewidth{0pt}
\tablehead{
\colhead{Molecule}&
\colhead{Transition} &
\colhead{$\nu$$_{0}$}&
\colhead{Telescope}&
\colhead{$\delta v$\tablenotemark{a}}&
\colhead{T$_{\rm sys}$\tablenotemark{b}}&
\colhead{HPBW\tablenotemark{c}}&
\colhead{$\eta_{\rm MB}$\tablenotemark{d}}&
\colhead{$T_{\rm MB}$\tablenotemark{e}}&
\colhead{Averaging\tablenotemark{f}}&
\colhead{$\Delta v$\tablenotemark{g}}&
\colhead{$\int{T_{\rm MB} dv}$\tablenotemark{h}} \\
 & &
\colhead{(MHz)}&
&
\colhead{(km s$^{-1}$)}&
\colhead{(K)}&
\colhead{($''$)}&
&
\colhead{(K)}&
\colhead{area ($\arcmin$~$\times$~$\arcmin$)}&
\colhead{(km s$^{-1}$)}&
\colhead{(K km s$^{-1}$)} 
}
\startdata
& & & & & & &  & & & & \\
& & & & & CB 65&  & & & & \\
& & & & & &  & & & & \\
\tableline

       CO        &$J=1\rightarrow0$                   &115271.2018     &IRAM-30m       &2.60   &400  &21 &0.73         &$<$0.03   &(0,0)    &\nodata  &\nodata  \\
                 &$J=2\rightarrow1$                   &230538.0000     &IRAM-30m       &1.30   &600  &11 &0.52         &$<$0.3    &(0,0)      &\nodata  &\nodata   \\
\tableline                                                                                                                              
& & & & & &  & & & & \\                                                    
& & & & & CB 101&  & & & & \\                                              
& & & & & &  & & & & \\                                                    
\tableline                                                                 
       $^{13}$CO &$J=1\rightarrow0$                   &110201.3541     &IRAM-30m       &0.05   &270  &22 &0.74    &1.51$\pm$0.23 &(0,0)  &0.44$\pm$0.04  &0.61$\pm$0.16\\
                 &$J=2\rightarrow1$                   &220398.6765     &IRAM-30m       &0.05   &2700 &11 &0.54    &1.1$\pm$0.5   &2$\times$2 &0.24$\pm$0.09  &0.27$\pm$0.21\\
       C$^{18}$O &$J=1\rightarrow0$                   &109782.1734     &IRAM-30m       &0.05   &290  &22 &0.74    &$<$0.10       &2$\times$2 &\nodata  &\nodata      \\
                 &$J=2\rightarrow1$                   &219560.3568     &IRAM-30m       &0.05   &1300 &11 &0.54    &$<$0.4        &2$\times$2 &\nodata  &\nodata      \\
       CO        &$J=1\rightarrow0$                   &115271.2018     &IRAM-30m       &0.05   &400  &21 &0.73    &6.3$\pm$0.6   &(0,0)  &0.86$\pm$0.03  &5.7$\pm$0.7  \\
                 &$J=2\rightarrow1$                   &230538.0000     &IRAM-30m       &0.05   &600  &11 &0.52    &6.6$\pm$1.0   &(0,0)   &0.95$\pm$0.05  &6.4$\pm$1.5  \\
       SiO       &$J=2\rightarrow1$                   &86846.9600      &IRAM-30m       &0.07   &700  &28 &0.77    &$<$0.7    &(0,0)     &\nodata  &\nodata      \\
       CS        &$J=2\rightarrow1$                   &97980.9500      &IRAM-30m       &0.06   &270  &25 &0.76    &$<$0.11   &2$\times$2     &\nodata  &\nodata      \\
                 &$J=5\rightarrow4$                   &244935.6435     &IRAM-30m       &1.22   &500  &10 &0.49    &$<$0.4    &(0,0)     &\nodata  &\nodata      \\
       CCS       &$J_{N}=2_{1}\rightarrow1_{0}$ &22344.0330      &Robledo-70m    &0.05   &60   &41 &0.38    &$<$0.5    &(0,0)    &\nodata  &\nodata      \\
       NH$_{3}$  &(J,K)=(1,1)             &23694.4955      &Robledo-70m    &0.50   &90   &39 &0.35    &$<$0.14   &1.7$\times$1.7    &\nodata  &\nodata      \\
\enddata
\tablenotetext{a}{Velocity resolution for each transition.}
\tablenotetext{b}{Typical system temperature.}
\tablenotetext{c}{Half power beam width of the telescope.}
\tablenotetext{d}{Main beam efficiency.}
\tablenotetext{e}{Peak main beam brightness temperature of the line. Uncertainties are
  2 $\sigma$. Upper limits are 3 $\sigma$.}
\tablenotetext{f}{Area over which data were averaged to obtain the
  quoted $T_{\rm MB}$,  centered on IRAS 16277$-$2332 for CB
  65 and on IRAS 17503$-$0833 for CB 101. (0,0) means that only the central spectrum was used.}
\tablenotetext{g}{$\rm Line$ width at half maximum, obtained from a Gaussian fit to the line. Uncertainties are 2 $\sigma$, and represent the error in the Gaussian fit.}
\tablenotetext{h}{Velocity integrated mean brightness temperature of the line. Uncertainties are 2 $\sigma$.}
\end{deluxetable}

\begin{deluxetable}{lr}
\tabletypesize{\scriptsize}
\tablecaption{Infrared data for IRAS 17503$-$0833 \label{tbl-cb101_ir}}
\tablewidth{0pt}
\tablehead{
\colhead{$\lambda$}&
\colhead{Flux density}
\\
\colhead{($\mu m$)}&
\colhead{(Jy)}

}
\startdata
1.25&   0.48 \\  
1.65&   0.82\\ 
2.17&   0.82\\
12 &    0.42 \\
25&     $<$0.28\\
60&     $<$0.42\\

\enddata

\end{deluxetable}

\end{document}